%% file: ASE.tex
\documentclass[10pt,conference]{IEEEtran}
\IEEEoverridecommandlockouts
\input{macro}

\usepackage{cite}
\usepackage{listings}
\usepackage{xcolor}
\usepackage{graphicx}
\usepackage{csquotes}
\usepackage[ruled, vlined]{algorithm2e}
\usepackage{circledtext}
\usepackage{pifont}
\usepackage{amsmath}
\usepackage{siunitx}
\usepackage{soul}
\usepackage{subcaption}
\usepackage[symbol]{footmisc}
\usepackage[]{collab}

\newcommand\alias{ErrorPrism\xspace}
\newcommand\company{ByteDance\xspace}

\collabAuthor{zb}{teal}{Zhuangbin}

\begin{document}

\thispagestyle{plain}
\pagestyle{plain}

\title{\alias: Reconstructing Error Propagation Paths in Cloud Service Systems}

\author{
    \IEEEauthorblockN{
        Junsong Pu$^{\ast}$, 
        Yichen Li$^{\S}$, 
        Zhuangbin Chen$^{\ast\dagger}\thanks{$^{\dag}$Zhuangbin Chen is the corresponding author.}$, 
        Jinyang Liu$^{\P}$, \\
        Zhihan Jiang$^{\ddag}$, 
        Jianjun Chen$^{\P}$, 
        Rui Shi$^{\S}$, 
        Zibin Zheng$^{\ast}$, 
        Tieying Zhang$^{\P}$
    }
    \IEEEauthorblockA{
        $^{\ast}$Sun Yat-sen University, Zhuhai, China,\ pujs@mail2.sysu.edu.cn, \{chenzhb36, zhzibin\}@mail.sysu.edu.cn \\
        $^{\ddag}$The Chinese University of Hong Kong, Hong Kong, China, zhjiang22@cse.cuhk.edu.hk \\
        $^{\P}$ByteDance US, San Jose, USA, \{jinyang.liu, jianjun.chen, tieying.zhang\}@bytedance.com \\
        $^{\S}$ByteDance, Beijing, China, \{liyichen.325, shirui\}@bytedance.com
    }
}


\maketitle

\input{Sections/00-abstract}


\input{Sections/01-introduction}
\input{Sections/02-background}
\input{Sections/04-method}

\input{Sections/06-deployed}
\input{Sections/07-related}
\input{Sections/08-conclusion}

\section*{Acknowledgement}

This work is supported by the National Natural Science Foundation of China (No. 62402536).

\balance
\bibliographystyle{IEEEtran}
\bibliography{bibliography}

\end{document}

%% file: macro.tex
\usepackage[]{collab}
\usepackage{enumitem}
\usepackage{balance}
\usepackage{xcolor}
\usepackage{setspace}
\usepackage{multirow}
\usepackage{multicol}
\usepackage{tabularx}
\usepackage{siunitx}
\usepackage{float}
\usepackage{color}
\usepackage{hyperref}
\usepackage{diagbox}
\usepackage{caption}
\usepackage{makecell}
\usepackage{colortbl}
\usepackage{tcolorbox}
\usepackage{mdframed}
\usepackage{listings}
\usepackage{booktabs}
\usepackage[utf8]{inputenc}
\usepackage{calligra}
\usepackage[normalem]{ulem}
\usepackage{indentfirst}
\usepackage{url}
\usepackage{soul}
\usepackage{nicematrix}
\usepackage{fontawesome5}
\usepackage{textcomp}
\usepackage{stfloats}
\usepackage{verbatim}
\usepackage{graphicx}
\usepackage{amsmath,amsfonts}
\usepackage[linesnumbered,ruled,vlined,noend]{algorithm2e}
\usepackage{array}
\usepackage[utf8]{inputenc}

\collabAuthor{zh}{purple}{Zhihan Jiang}
\collabAuthor{yc}{blue}{Yichen Li}
\collabAuthor{js}{cyan}{Junsong Pu}

\newcommand{\boxmargin}{1mm}

\newtcolorbox{myboxa}[2][]{
    colback=gray!10!white,
    colframe=black, enhanced,
    attach boxed title to top left={yshift=-2mm,xshift=5mm},
    title=#2,#1
}
\newtcolorbox{myboxb}[2][]{
    boxsep=3pt,
    left = \boxmargin, right = \boxmargin, top = \boxmargin, bottom = \boxmargin,
    title={#2},#1
}
\newtcolorbox{myboxc}{
    colback=gray!15!white,
    arc = 0pt, outer arc = 0pt,
    boxsep=0pt, left = 3pt, right = 0pt, top = 0pt, bottom = 0pt, 
    leftrule=3pt, bottomrule=0pt,toprule=0pt, rightrule=0pt,
    left = \boxmargin, right = \boxmargin, top = \boxmargin, bottom = \boxmargin
}
\newtcolorbox{myboxd}{
    colback=gray!10,
    colframe=black,
    width=\columnwidth,
    arc=1mm, auto outer arc,
    boxrule=0.5pt,
}

\definecolor{myyellow}{HTML}{FFF2CC}
\newcounter{finding}

\definecolor{myyellow}{HTML}{FFF2CC}
\newcounter{insight}

\newcounter{challenge}

\definecolor{mygreen}{HTML}{AFCFA5}
\newcounter{opportunity}

%% file: Sections/00-abstract.tex
\begin{abstract}

Reliability management in cloud service systems is challenging due to the cascading effect of failures. Error wrapping, a practice prevalent in modern microservice development, enriches errors with context at each layer of the function call stack, constructing an error chain that describes a failure from its technical origin to its business impact.
However, this also presents a significant traceability problem when recovering the complete error propagation path from the final log message back to its source.
Existing approaches are ineffective at addressing this problem.
To fill this gap, we present \alias in this work for automated reconstruction of error propagation paths in production microservice systems.
\alias first performs static analysis on service code repositories to build a function call graph and map log strings to relevant candidate functions.
This significantly reduces the path search space for subsequent analysis. Then, \alias employs an LLM agent to perform an iterative backward search to accurately reconstruct the complete, multi-hop error path.
Evaluated on 67 production microservices at ByteDance, \alias achieves 97.0\% accuracy in reconstructing paths for 102 real-world errors, outperforming existing static analysis and LLM-based approaches.
\alias provides an effective and practical tool for root cause analysis in industrial microservice systems.
\end{abstract}

\begin{IEEEkeywords}
Cloud Computing, System Reliability, Root Cause Analysis, Error Tracking, Log Analysis
\end{IEEEkeywords}

%% file: Sections/01-introduction.tex
\section{introduction}
\label{sec:intro}

The microservice architecture has become the dominant paradigm for building complex and large-scale cloud service systems. While this architectural style enhances scalability, it fundamentally complicates system observability and diagnostics~\cite{DBLP:conf/sigsoft/ChenKLZZXZYSXDG20,DBLP:conf/icse/DangLH19,DBLP:conf/cikm/WangLZZWYF0W24,DBLP:journals/corr/abs-2406-11213}. A single end-user request can trigger a cascade of invocations across a distributed graph of services, making root cause analysis (RCA) significantly more challenging than in a monolithic environment. Consequently, a failure in one service can manifest as a symptom in another, requiring a holistic view to trace the fault back to its origin.

The error handling paradigm of a programming language plays a fundamental role in the process of failure diagnosis~\cite{DBLP:conf/nsdi/Li0NPS24}.
For example, Java and Python have an exception system which interrupts the regular execution flow of programs. When an exception is thrown, it automatically propagates up the function call stack unless it is explicitly caught and handled.
However, many languages prevalent in modern microservice development (e.g., Go, Rust) advocate for a different error handling principle, which treats errors as explicit return values.
This non-disruptive approach requires developers to explicitly handle potential failures at each call site.
As an error is passed up the call stack, each layer can programmatically ``wrap'' it with its own context, constructing a layered error chain that precisely describes a failure from its technical origin to its business impact.
This practice is also common in large-scale open-source systems. Our analysis of the Kubernetes repository~\cite{google_kubernetes} uncovers 18,762 instances of error wrapping, which far outnumbers the 4,511 error and fatal log statements.

From a developer's perspective, this error wrapping at each layer enables the construction of a self-contained diagnostic artifact that describes how a low-level fault escalates into a high-level failure.
However, when the wrapped error is logged for diagnosis, its entire hierarchical structure is typically flattened into a single log string.
From a site reliability engineer's (SRE's) perspective, this introduces a significant traceability challenge when recovering the complete error propagation path from the symptom (i.e., the composite log message) to its origin.
The log itself contains no explicit pointers that attribute the different semantic fragments back to their specific source code locations, leading to multiple potential backward paths.
We term this problem \textit{Error Obfuscation}.
This ambiguity arises from several practical factors. For example, a generic error string like ``operation failed'' may be used in dozens of unrelated functions.
These issues present significant operational challenges in \company. As a world-leading cloud service provider, Bytedance operates thousands of microservice applications.
When failures occur, SREs need to resort to complex, manual investigations, inspecting the source code to retrieve the error propagation path.

Existing approaches are unable to effectively address the error obfuscation problem.
For example, log-based methods~\cite{DBLP:conf/kdd/NandiMADB16,DBLP:conf/ccs/Du0ZS17,DBLP:conf/osdi/ZhaoRLYS16,DBLP:conf/issta/JiangL00HGCZL24}, which often rely on log parsing, are fundamentally inapplicable. They operate on the assumption of a one-to-one mapping between a logging statement and its template. This does not hold as error wrapping allows a single logging point to produce multiple log events depending on the underlying error propagation path.
Moreover, errors are typically logged only at the end of the propagation chain, instead of every intermediate function.
This makes them insufficient for recovering the entire program execution flow.
In addition, logs alone may lack the detailed context needed to deterministically trace a failure back to its origin.
The source code, in contrast, explicitly defines the control flow, error handling logic, and the execution context that describe how an error is enriched and passed between functions.
However, traditional static analysis is unable to fully harness this information.
The prevalence of asynchronous operations and inter-process communication creates an explosion of potential execution paths that are difficult to model statically.
It also lacks the ability to comprehend code semantics essential for solving path ambiguity.



To bridge this gap, we propose \alias, a framework for automated reconstruction of error propagation paths in production microservice systems.
\alias employs a hybrid methodology that integrates the structural precision of static analysis with the semantic reasoning capabilities of Large Language Models (LLMs).
Specifically, \alias performs static analysis on service code repositories to construct a function call graph and index error-related string constants with their function-level provenance.
This step maps potential error fragments to relevant candidate functions, significantly pruning the search space.
Then, an LLM-guided agent performs an iterative, backward search to trace the error log to its origin.
Based on the precomputed artifacts and source code, the agent jointly reasons about control flow, error-wrapping patterns, and semantic context to resolve ambiguities for path reconstruction.
We have deployed \alias in our production environment at \company, where it monitors a suite of cloud infrastructure services. In an evaluation on 102 real-world errors, \alias achieves an overall accuracy of 97.0\% in reconstructing the complete error propagation path. This performance outperforms both pure static analysis and naive LLM-based approaches, providing a practical and effective tool for automating RCA in modern cloud systems.

In summary, our major contributions are as follows:

\begin{itemize}[noitemsep,leftmargin=5.5mm]
    \item We identify and formalize the problem of \textit{error obfuscation}, a critical traceability challenge in modern cloud services where the common practice of error wrapping leads to ambiguous log messages that hide an error's true propagation path.
    \item We design and implement \alias to address the error obfuscation problem. \alias uses static analysis to dramatically prune the search space and then leverages the LLM's semantic reasoning to perform an iterative backward search for accurate error path reconstruction.
    \item We evaluate \alias on a large-scale production cloud platform at ByteDance. Our results show that \alias successfully reconstructs the propagation path for 97.0\% of 102 real-world errors, significantly outperforming existing approaches and demonstrating its practical effectiveness.
\end{itemize}




%% file: Sections/02-background.tex
\section{Background and Problem Statement}
\label{sec:background}

\subsection{Error Handling in Modern Microservice Systems}
\label{sec:error_handling}

In microservice applications (especially those written in Java and Python), faults are typically signaled by throwing exceptions. An exception is a disruptive event that diverts the program from its normal execution path, propagating up the call stack until it is caught by a designated handler.
Upon catching a failure, this handler logs critical diagnostic information, such as the exception's message and a detailed stack trace, which serves as the primary artifact for RCA and debugging.
In this process, SREs often need to navigate a vast collection of distributed logs from multiple services. It involves correlating log entries using tracing IDs~\cite{DBLP:conf/nsdi/ZhangXAVM23,10.1145/3728888}, timestamps, and other metadata to manually reconstruct the sequence of events and pinpoint the original source of the failure. This log analysis process is often complex, time-consuming, and requires deep system knowledge~\cite{DBLP:conf/icse/SahaH22,DBLP:conf/icse/ShettyBKRN021,huang2024faultprofit}.

To mitigate this inherent complexity, many programming languages prevalent in modern microservice development (e.g., Go~\cite{gerrand2011error}, Rust~\cite{kas2024static}) advocate for an error handling principle that treats errors as explicit return values~\cite{kas2024static,errorcode_detect}.
In this paradigm, a function that may fail will return its result encapsulated in a type that represents both success and failure, e.g., Rust's \texttt{Result$<$T, E$>$} enum and Go's \texttt{(T, error)} multi-value return.
This error handling philosophy offers several advantages that make it particularly well-suited for building robust services:

\begin{itemize}[noitemsep,leftmargin=5.5mm]
    \item \textit{First, it does not interrupt the normal execution flow of the program}. As errors are returned as a regular value, developers are compelled to explicitly handle potential failures at each step, preventing unhandled exceptions from unexpectedly crashing a service.

    \item \textit{Second, it enables compile-time correctness guarantees.} The explicit nature of error-return types allows static analysis tools and compilers to verify that all possible error paths are handled. This shifts error management from an error-prone runtime discipline to a compile-time guarantee.
    

    \item \textit{Third, it makes errors transmissible as structured data}.
    As first-class values, errors can be seamlessly transmitted both within a service using concurrent mechanisms like thread-safe channels, and between services via RPC responses or message queues. This allows services to programmatically enrich errors with structured context for subsequent RCA.

\end{itemize}

\noindent A disciplined implementation of this paradigm involves a practice known as ``error wrapping'' or ``context enrichment.'' As an error value is passed up the propagation path (which may span the call stack, component layers, and asynchronous channels) from a low-level function to a high-level one, each intermediate layer can add its own contextual information. This process constructs a detailed \textit{error propagation chain}, which can precisely describe a failure with structured context, from its technical origin to its business impact.

\begin{figure*}[t]
\centering
\includegraphics[width=0.91\linewidth]{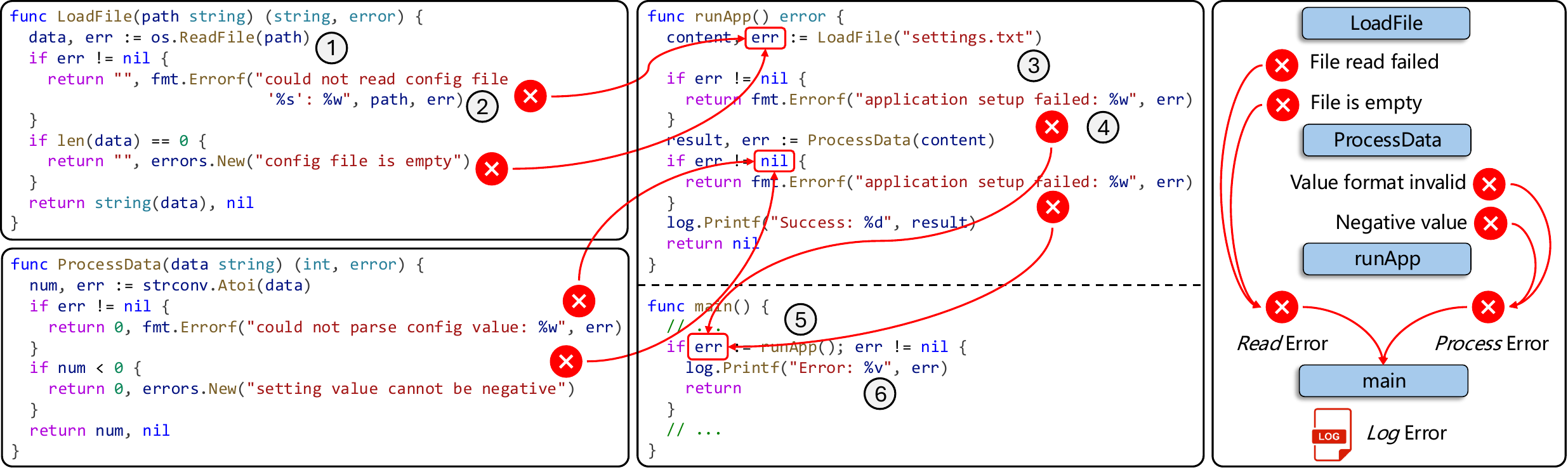}
\caption{An Example of Error Wrapping and Error Propagation}
\label{fig:error_propa_example}
\vspace{-9pt}
\end{figure*}

Fig.~\ref{fig:error_propa_example} presents an example in Go to illustrate the concepts described above, in which the program implements a task of loading and parsing a numerical setting from a configuration file. The \texttt{main} function orchestrates the process by calling \texttt{runApp}, which in turn delegates to two lower-level utilities: \texttt{LoadFile} for file system I/O and \texttt{ProcessData} for string parsing. 
In each function, errors are treated as explicit return values.
For instance, \texttt{LoadFile} is declared as \texttt{func LoadFile(path string) (string, error)}, explicitly stating that it will return either a \texttt{string} on success or an \texttt{error} on failure.
This forces the calling function, \texttt{runApp}, to handle the potential failure immediately with an \texttt{if err != nil} check.
Moreover, the implementation showcases the practice of error wrapping.
Each function in the call stack adds a layer of context that is specific to its own level of abstraction and responsibility.
This is achieved in Go using the \texttt{\%w format} verb within \texttt{fmt.Errorf}, which creates a new error that contains the original error.
To see this in action, we trace the execution flow for the failure where the \texttt{settings.txt} file does not exist.

\begin{enumerate}[noitemsep,leftmargin=5.5mm]
    \item \texttt{ReadFile} cannot find the file and returns an error: \texttt{open settings.txt: no such file or directory} (\ding{192}). This is the technical root cause.
    However, on its own, this low-level OS error is of limited utility for quick diagnosis. It is precise about what happened at the system level, but it provides no application-level context about why that file was being accessed. An operator seeing such a log would have to manually investigate which part of the application needs this file and how critical it is.

    \item \texttt{LoadFile} catches this generic OS error and wraps it with its specific task's context: \texttt{fmt.Errorf("could not read config file '\%s': \%w", path, err)} (\ding{193}$\rightarrow$\ding{194}).
    This specifies the file's role in the application, i.e., a configuration file, which immediately narrows down the failure scope.

    \item The resulting error is passed up to the \texttt{runApp} function, which wraps it again, adding the highest-level context: \texttt{fmt.Errorf("application setup failed: \%w", err)} (\ding{195}$\rightarrow$\ding{196}). This final layer explains the ultimate business impact of the failure, i.e., the entire service could not initialize.

    \item Finally, the \texttt{main} function uses this detailed error object to produce a single, complete log message, which shows the full chain of wrapped errors (\ding{197}):
\end{enumerate}

\begin{lstlisting}
Error: application setup failed: could not read config file 'settings.txt': open settings.txt: no such file or directory
\end{lstlisting}

This layered error structure offers a fundamental advantage over a traditional stack trace. While a stack trace shows the code execution path, i.e., a sequence of function calls, it lacks semantic context. An engineer must manually inspect the code at each frame to infer the program's intent. In contrast, the composite error object constructs a causal chain of the failure, where each layer explicitly states its contribution to the overall operation. This process transforms low-level signals into a rich, self-contained diagnostic artifact that is immediately actionable for both operators and developers.

However, the benefits of this error chain representation can be lost at the system's observability layer.
By serializing the entire error hierarchy into a single, flat string, the system produces logs that are human-readable but machine-unfriendly.
This introduces significant challenges for automated root cause analysis at the system level, as we discuss next.

\subsection{Error Obfuscation}
\label{sec:error_obfuscation}

A fundamental challenge in log-based failure diagnosis~\cite{DBLP:conf/kdd/NandiMADB16,DBLP:journals/infsof/YuanLSL20,DBLP:conf/asplos/YuanMXTZP10,DBLP:conf/sigsoft/Zhou0X0JLXH19,wang2024comprehensivesurveyrootcause,he2021survey,huang2024demystifying,jiang2025l4} is the reconstruction of the complete error propagation path, which traces the failure from its initial source to its final observable impact.
In the aforementioned error handling mechanism, the final error log is assembled dynamically across multiple functions.
Thus, a single top-level logging statement can produce messages with different underlying error chains depending on the failure's origin.
We term this phenomenon \textit{Error Obfuscation}.
This leads to several key problems in practical service reliability management:

\begin{itemize}[noitemsep,leftmargin=5.5mm]
    \item The flattening of rich error objects into simple text strings creates a significant traceability challenge. It is difficult to attribute the different semantic parts of a composite log message back to the disparate source code locations.
    In reality, an error path is rarely a simple, in-process call stack. Instead, it frequently crosses asynchronous boundaries where traditional stack tracing fails and the causality is obscured. Moreover, it often traverses service boundaries via RPC calls, leaving the local service with only a generic network error that masks the true root cause.

    \item Error obfuscation renders many existing log-based analysis techniques inapplicable~\cite{DBLP:conf/ccs/Du0ZS17,24WittkoppLogRCA,li2025coca,DBLP:journals/corr/abs-2107-05908,DBLP:conf/issre/HeZHL16,liu2023scalable}. These approaches, particularly log parsing~\cite{jiang2024lilac,huang2025no}, often assume a one-to-one relationship between a logging statement and the log template it produces. However, with error wrapping, a single logging point can produce a multitude of distinct log events.
    To illustrate, consider a different failure within Fig.~\ref{fig:error_propa_example}, where \texttt{settings.txt} contains non-numerical data. In this case, the same logging statement in \texttt{main} produces an entirely different log message:

\begin{lstlisting}
Error: application setup failed: could not parse config value: strconv.Atoi: parsing "invalid": invalid syntax
\end{lstlisting}

    This variability breaks the fundamental assumption of template-based log analysis ~\cite{DBLP:journals/corr/abs-2107-05908,DBLP:conf/issre/HeZHL16,DBLP:conf/icse/ZhuHLHXZL19}.

    \item The utility of error wrapping is fundamentally constrained by inconsistent developer practices. This manifests in two opposing failure modes: \textit{under-enrichment}, where developers forget to wrap errors or omit crucial context, and \textit{over-enrichment}, where logs contain verbose technical details incomprehensible to external consumers like system operators.
    Without a clear standard governing what context should be preserved, a semantic drift emerges between the internal error structure and its final logged output.

\end{itemize}




Given these problems, a natural question arises: \textit{why not simply log the error in every function along its propagation path?} While seemingly straightforward, this approach is counter-productive for several reasons. First, it generates extreme log verbosity, overwhelming observability systems and creating an unmanageable signal-to-noise ratio. Second, and more importantly, it results in contextual fragmentation. Each log entry contains only the information available at its specific layer, forcing engineers to manually correlate multiple log lines to reconstruct the full path. This is precisely the burden that structured error wrapping aims to eliminate~\cite{google_flogger_best_practice}.

\subsection{Problem Statement}

The issue of error obfuscation renders a significant observability gap between an error's final log message and its original root cause.
To bridge this gap, we perform \textit{error propagation tracking} in this paper, which is to automatically reconstruct the causal chain of an error as it traverses function calls, asynchronous boundaries, and service borders.

Particularly, the input to this problem is an error log $L$, which is a composite string representing a specific type of failure, and the source code repositories $C$ of the microservice application under study. The output goal is to find the error propagation path, which is an ordered sequence of functions $P = <f_n, f_{n-1}, ..., f_1>$. In this sequence, the first function $f_n$ is the one that ultimately prints the error log $L$, and $f_1$ is the source function where the error originates. For any two adjacent functions in the path, the latter passes the error to the former via a direct or indirect calling relationship, thus forming a complete propagation chain from the error's origin, $f_1$, to the final logging point, $f_n$.

%% file: Sections/04-method.tex
\section{Methodology}
\label{sec:method}

\begin{figure*}[t]
    \centering
    \includegraphics[width=0.88\linewidth]{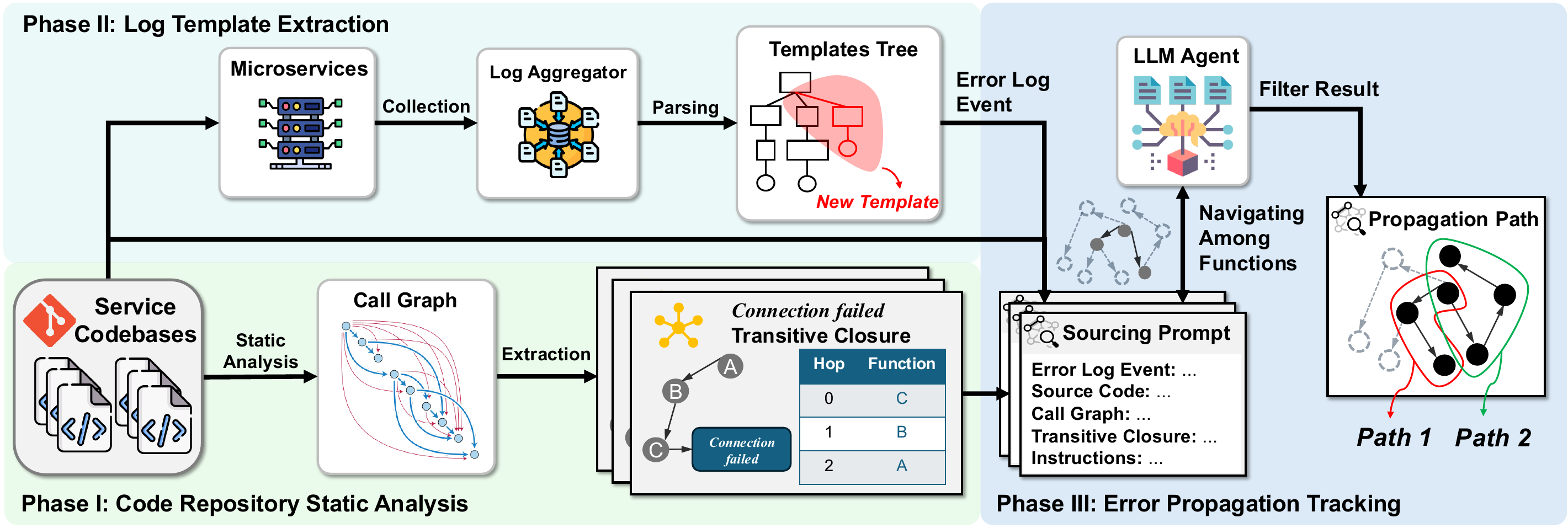}
    \caption{The Overview Framework of \alias}
    \label{fig:framework}
    \vspace{-10pt}
\end{figure*}

\subsection{Overview}

In this section, we present the design of \alias.
The overall framework is shown in Fig.~\ref{fig:framework}, which consists of three phases: \textit{code repository static analysis}, \textit{log template extraction}, and \textit{error propagation tracking}.
The first phase constructs a function call graph based on static analysis. This pre-computation creates a reverse index that can rapidly map a runtime error log to a small set of relevant candidate functions, significantly narrowing the search space.
The second phase clusters and templates raw logs. This process distills the high volume of production logs into different log events, each of which corresponds to a unique error propagation path.
The last phase employs an LLM-guided agent in an iterative search to reconstruct the failure path. Using the focused context from the previous phases, the agent reasons about the code to trace the error backward from the error log to its origin, progressively building the complete, multi-hop propagation path.

\subsection{Code Repository Static Analysis}
\label{sec:phase1-static_analysis}

Our approach begins with an offline static analysis performed on the microservice code repositories, which consists of three sequential steps, i.e., function call graph construction, error-related string constant extraction, and string reachability computation across the call graph.
This pre-computation creates a reverse index that drastically narrows the search space for the LLM, providing a focused set of candidate functions for it to analyze when tracing the error's execution path.

A function call graph (FCG) is a directed graph, denoted $\mathcal{G} = (\mathcal{F}, \mathcal{E})$. 
Each vertex $f \in \mathcal{F}$ represents a function within the service's codebase, and a directed edge $(f_i, f_j) \in \mathcal{E}$ exists if function $f_i$ contains a call to function $f_j$.
In our implementation, we use an internally maintained tool that parses the source code to identify all function definitions and invocation sites. We utilize the relatively fast Rapid Type Analysis algorithm to construct the call graph, and its false-positive edges will be pruned by subsequent methods.

With the call graph established, the next step is to identify and associate potential error-message fragments with the functions that introduce them. We perform a targeted scan of the golang's SSA representation for each function to extract all string constants.
This involves an intra-procedural data-flow analysis to collect string constants that are directly or indirectly referenced by logging statements (e.g., \texttt{logger.Error}) and error-creation functions (e.g., \texttt{errors.New}, \texttt{fmt.Errorf}).
This process yields a mapping, $\sigma : \mathcal{F} \to 2^S$, from each function $f$ to the set of relevant string constants $S$ that it directly references.
In Fig.~\ref{fig:error_propa_example}, this step would create the direct associations of: $\sigma$\texttt{(main)=\{"Error: \%v"\}}
\noindent $\sigma$\texttt{(LoadFile)=\{"could not read config file '\%s': \%w", "config file is empty"\}}, etc.


The final step is to compute the reachability of these string constants throughout the call graph. A single log message often contains a composite string built from fragments contributed by multiple functions in a call chain. To trace such a message, we must know not only which strings a function references directly, but also which strings it can indirectly reach from the functions it calls.
We formalize this concept as the \textit{constant transitive closure}, denoted $\mathcal{C}_k(f)$, which represents the set of all string constants reachable within a call depth of at most $k$ from a function.
A string directly referenced by $f$ is at a call depth of 0, while a string referenced by a function that $f$ calls is at a call depth of 1, and so on. The closure is defined recursively:

\begin{equation}
    \mathcal{C}_k(f) = 
    \begin{cases}
    \sigma(f) & \text{if } k = 0 \\[8pt]
    \sigma(f) \cup \bigcup\limits_{(f,g) \in \mathcal{E}} \mathcal{C}_{k-1}(g) & \text{if } k \geq 1
    \end{cases}
\end{equation}

\noindent In our implementation, we set a call depth limit of $k=3$ to balance effectiveness and computational cost.
To compute this finite closure efficiently for all functions, we employ a backward Breadth-First Search (BFS) starting from each string constant. For each string, the BFS traverses the call graph in reverse, propagating the string's reachability backward to its callers up to the three-hop limit. The final output of this phase is a pre-computed index that, for any function, provides a lookup of all string fragments it could potentially contribute to a log message, setting the stage for the dynamic log analysis.

\subsection{Log Template Extraction}
\label{subsec:template_extract}

The second phase addresses the immense volume and variability of logs generated by production systems. Given that a service can produce billions of log entries daily, performing deep source code analysis on each individual message is computationally infeasible~\cite{DBLP:conf/kbse/ChenLSZWLYL21,DBLP:conf/icse/Zhao0PWWZCZNWWZ20,DBLP:journals/tkdd/WenCZSXLQMT24,10.1145/3611643.3613078}. Our approach, therefore, is to first distill this raw data stream into a concise set of unique and actionable error patterns. This is achieved through a multi-stage process of log clustering and templating.

To start, we perform a coarse-grained clustering by bucketing logs based on their static source code origin (i.e., file name and line number), which is often included as metadata in structured logs.
Logs in each cluster all originate from the same logging statement.
Next, we apply log templating within each coarse-grained group to convert raw log messages into structured templates (or events) by separating the static text from variable parameters.
For this, we employ Drain3, an efficient, streaming-capable log parsing approach.
This templating step is fundamental to our methodology for two critical reasons.
First, it isolates the dynamic parameters of the log messages.
They are runtime variables that do not appear as constants in logging statements, which makes them untraceable with static analysis techniques.
Second, and more critically, it mitigates the ambiguity caused by error obfuscation. 
As discussed in Sec.~\ref{sec:error_obfuscation}, a single top-level logging statement can produce different log events depending on the underlying error chains.
Log templating helps distinguish error propagation paths, i.e., each log template corresponds to one error chain.
During runtime analysis, \alias focuses specifically on logs with an \texttt{error} severity level. To avoid redundant computation, we maintain a historical repository of previously processed error logs and their corresponding propagation paths.

\subsection{Error Propagation Tracking}


Given an input log event, in this phase we reconstruct its error propagation path based on the function call graph $\mathcal{G}$ and constant transitive closure $\mathcal{C}_k(f)$. Intuitively, this can be done by first identifying the log-generating function (Sec.~\ref{subsec:template_extract}), and then recursively tracing through the call graph based on the string constants present in the error template.
However, we face two challenges that render purely static or string-matching techniques insufficient.
First, while $\mathcal{C}_k(f)$ can identify a set of candidate functions that contain relevant string fragments, this process often yields a large number of false positives. A generic phrase like ``operation failed'' could appear in dozens of unrelated functions across the codebase. This problem is compounded by developers' nonstandard logging practices (Sec.~\ref{sec:error_obfuscation}).
Identifying the true path requires understanding the specific inter-procedural control flow and error-handling logic within each function. Although multi-level call-site-sensitive pointer analysis could theoretically filter some of these invalid routes by tracking the flow of the error variable, its prohibitive computational overhead makes it impractical.
The second challenge is the presence of broken paths within the statically-constructed function call graph. Modern software relies heavily on dynamic dispatch mechanisms, such as RPC invocations and asynchronous messaging, which challenge static analysis. For example, an RPC call is typically represented in the static graph as a mere invocation of a generic library function (e.g., \texttt{client.Call}), with the actual business logic endpoint specified as a string parameter. Consequently, the static error propagation link is interrupted at this point.

To address these limitations, we leverage the semantic understanding and reasoning capabilities of LLMs. The LLM acts as an intelligent agent that guides the traversal of the function call graph. It goes beyond simple string matching by analyzing the full context of the source code. To prune false positives, the LLM examines the control flow logic, the structure of error-wrapping calls (e.g., \texttt{fmt.Errorf}), and the semantic relevance of a candidate function to the overall error message. Moreover, to bridge the gaps in the static call graph, the LLM uses its code comprehension to interpret dynamic invocation patterns. It can parse the string arguments in an RPC call, identify the target function in a different service, and intelligently resume the tracing process from that point. 

\subsubsection{Candidate Scoping and Indexing}

Before path exploration, we perform a crucial pre-processing step to reduce the search space and index the relevant data. The objective is to filter developer-specified microservice repositories down to a small, relevant set of candidate functions.

This process starts by identifying functions that contain string constants semantically related to the target error log event.
For format strings, we parse them into their static text fragments, e.g., \texttt{"receive package \%d from source \%s"} is parsed into the fragments [\texttt{"receive package"}, \texttt{"from source"}]. We define the matching rule for a format string and a log template: a format string is considered a match if and only if all of its static text fragments are found within the log template. A function is then selected as a candidate if it contains at least one matching string, regardless of whether it is a format string or a literal string.
Once the candidate functions are identified, we build a comprehensive index, which maps each candidate function's identifier to its essential metadata, including its file path and complete source code. This stage transforms the raw code repositories into a structured, self-contained map, providing the main exploration algorithm with immediate, efficient access to all necessary information.

\begin{figure}[t]
    \centering
    \begin{subfigure}[b]{\linewidth}
        \centering
        \includegraphics[width=0.88\linewidth]{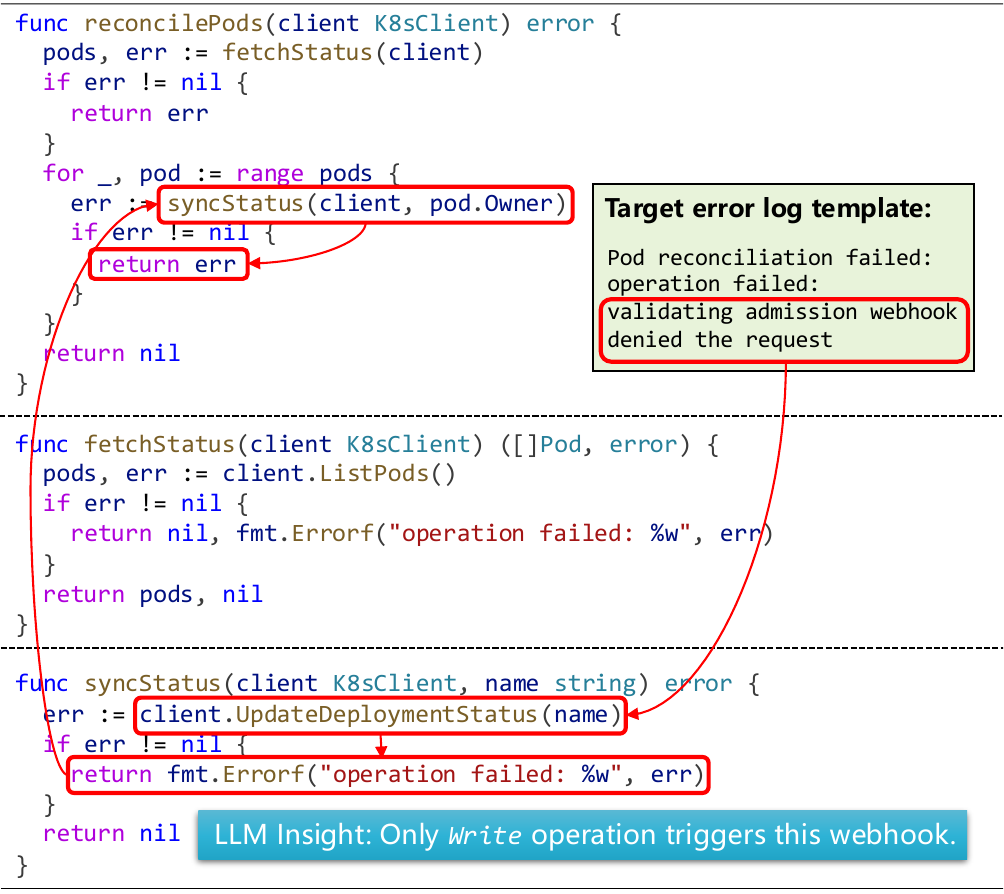}
        \caption{Next-hop Selection in \alias}
        \label{fig:next_hop_selection}
    \end{subfigure}
    
    \vspace{0.2cm}
    
    \begin{subfigure}[b]{\linewidth}
        \centering
        \includegraphics[width=0.86\linewidth]{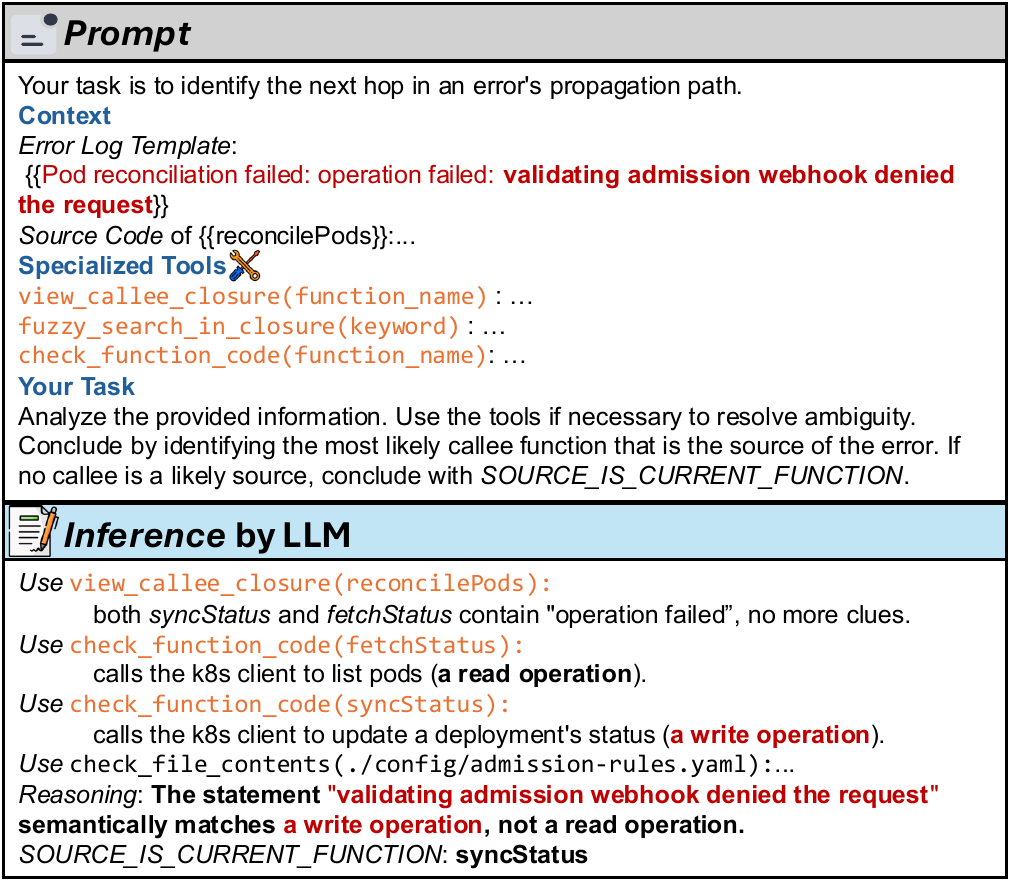}
        \caption{Example Prompt in \alias}
        \label{fig:llm_prompt}
    \end{subfigure}
    
    \caption{LLM-guided Iterative Propagation Path Construction}
    \label{fig:llm_workload}
    \vspace{-10pt}
\end{figure}


\subsubsection{LLM-guided Path Reconstruction}

While the static analysis phase effectively prunes the search space, it is inherently limited by semantic ambiguity. To overcome this, \alias employs an iterative reconstruction process guided by a LLM, which is configured as an autonomous agent. Using the ReAct framework~\cite{DBLP:conf/iclr/YaoZYDSN023,DBLP:conf/sigsoft/RoyZBBLFR24}, the agent mimics an expert SRE's diagnostic process, combining reasoning with tool use to trace an error's path backward. This entire exploration is orchestrated as a Breadth-First Search (BFS) over the call graph, ensuring a systematic and complete analysis of all potential propagation paths. In \company practice, logging middleware often includes the source function and line number for each log entry. This allows us to use the specified function as a precise starting point for our traversal.
When this information is unavailable, we could use constant propagation analysis to find starting functions by matching their logged string constants with the target log's prefix.
The agent's primary task is to, given a function in the error path, identify the specific upstream callee responsible for the error. To do this, 
it is equipped with a specialized toolset that allows it to query the static analysis artifacts and source code on demand.

\begin{itemize}[noitemsep,leftmargin=5.5mm]
    \item \texttt{view\_callee\_closure(function)}: This tool queries the pre-computed constant transitive closure. It serves as a rapid, first-pass filter, allowing the agent to check which of a function's callees are statically associated with error strings found in the log.

    \item \texttt{check\_function\_code(function)}: This tool retrieves the full source code of a specified function. It is essential for deep semantic analysis when string matching is insufficient, enabling the agent to reason about business logic, code comments, and overall function intent.

    \item \texttt{fuzzy\_search\_in\_closure(keyword)}: This tool performs a fuzzy search for a keyword within the string constants of all functions. It is designed specifically to bridge broken paths in the call graph, such as by identifying an RPC endpoint defined as a string literal.
\end{itemize}

We walk through the agent's workflow using the example in Fig.~\ref{fig:llm_workload}, tracing error log \texttt{"Pod reconciliation failed: operation failed: validating admission webhook denied the request"}. The process begins at the \texttt{reconcilePods} function.
First, the agent confronts the ambiguous error fragment \texttt{"operation failed"}. It uses \texttt{view\_callee\_closure} and confirms that both \texttt{fetchStatus} and \texttt{syncStatus} are potential candidates, as both can produce this generic error. At this point, static analysis hits a wall.
Faced with this ambiguity, the agent pivots to a deeper semantic analysis. It uses \texttt{check\_function\_code} to inspect the source code of both candidates. This reveals a critical distinction, i.e., \texttt{fetchStatus} performs a read operation by listing Kubernetes resources, while \texttt{syncStatus} performs a write operation by updating a resource's status.
The agent synthesizes the code-level distinction with the semantic content of the error log: \texttt{"validating admission webhook denied the request"}. The agent deduces that validating admission webhooks in Kubernetes intercept write operations (like creating or updating resources), not read operations. To further verify this deduction, it inspects the webhook's configuration files. This correlation between the denied write operation and the code allows it to identify \texttt{syncStatus} as more likely the true error source.
Once \texttt{syncStatus} is confirmed as the next hop, it is added to the BFS queue for the subsequent iteration. The agent will then be reinvoked on \texttt{syncStatus} to trace the path further upstream. This iterative process continues until the agent determines that a path's ultimate origin is found
or the BFS queue is empty, signifying that all potential error propagation paths have been fully explored.

%% file: Sections/06-deployed.tex
\section{Evaluation}
\label{sec:evaluation}

In this section, we evaluate the performance of \alias in our production environment. In particular, we aim to answer the following research questions:

\begin{itemize}
    \item \textbf{RQ1 (Effectiveness)}: How effective is \alias in reconstructing the propagation path of errors?

    \item \textbf{RQ2 (Efficiency)}: How efficient is \alias in terms of inference time?
\end{itemize}

The evaluation is conducted on a large-scale cloud service platform at \company. This platform, built on a microservice architecture, includes a suite of critical applications, such as billing systems, scheduling systems, and middleware modules.
Our study encompasses 67 representative microservices that operate on the platform.
We collect their source code repositories which are written in Go, totaling 988k lines of code.
Over time, the development team of this platform has accumulated a wealth of knowledge regarding historical service failures through their daily development and maintenance activities.
They maintain a detailed post-mortem report for each significant failures, which documents the complete failure investigation process based on different observability data, including logs, metrics, and distributed traces. Particularly, the report also details the manual source code analysis necessary for failure diagnosis.

\begin{figure}[t]
\centering
    \includegraphics[width=0.93\linewidth]{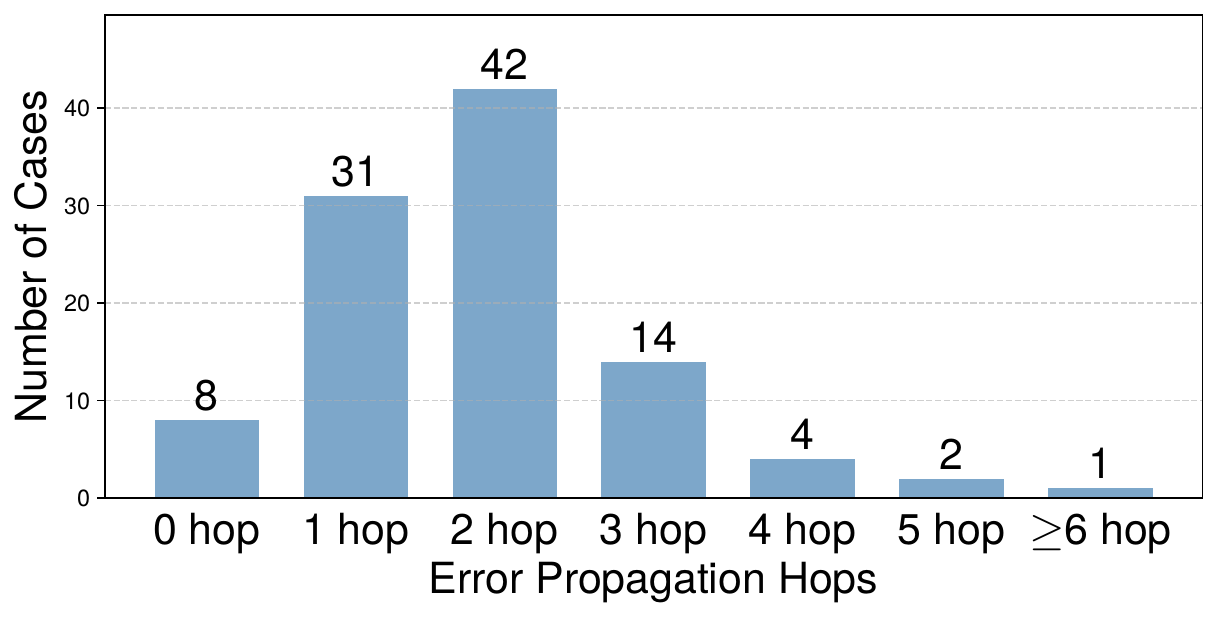}
\captionsetup{skip=0pt}
\caption{Distribution of error propagation hop}
\label{fig:hop_distribution}
\vspace{-15pt}
\end{figure}


\subsection{Evaluation Design}

\subsubsection{Dataset Construction}

The evaluation dataset is built through a multi-stage process to ensure its real-world representativeness and accuracy, including data collection, parsing, and ground-truth establishment.
We begin by collecting a corpus of over three million raw error logs associated with historical service failures.
To distill this large volume of unstructured text into structured events, we apply the Drain3 log parsing algorithm (Sec.~\ref{subsec:template_extract}). This step clusters the raw logs and parses them into 257 unique log templates, each representing a distinct type of error event observed in the system.
The ground truth for each template consists of its complete, manually verified error propagation path, i.e., the sequence of function calls, including those across asynchronous and service boundaries, from the error's origin to the final logging statement.
This is accomplished through a two-pronged approach. 
For a subset of these templates, the paths have been analyzed and recorded in detailed post-mortem reports, which are crucial for root cause analysis. For the remaining templates whose path is not available in the report, we collaborate with the development teams to construct it. This involves a meticulous process of manual source code analysis, where engineers trace each error backward from its logging statement through the complex microservice call chain to definitively identify its root cause.

The above process yields a final dataset of 102 distinct and representative error events, each paired with its ground-truth propagation path. To characterize the complexity of these real-world failures, Fig.~\ref{fig:hop_distribution} shows the distribution of their path lengths, measured in hop count, which is defined as the number of times an error is contextually wrapped during its propagation. The data reveal that the vast majority of errors (92.2\%) are not logged at their source, frequently requiring multiple hops to trace. Particularly, 20.6\% of these errors have a path with $\ge3$ hops.
This distribution underscores the necessity of a tool capable of automated propagation tracking.

\subsubsection{Baseline Methods}

We compare \alias against a static analysis approach and three LLM-based methods. These baselines are selected because they represent the state-of-the-art and key alternative strategies in both traditional program analysis and generative AI for code.


\textbf{Static Analysis.} This baseline is similar to the candidate generation phase of \alias (Sec.~\ref{sec:phase1-static_analysis}), where we construct a static call graph for the service.
Thus, it also serves as an ablation study of our method.
We enhance this baseline in two ways for a fair comparison. First, we integrate invocation data from the observability platform to add corresponding call edges to the static graph. Second, we employ a flow-sensitive, intra-procedural pointer analysis to prune impossible error propagation paths within individual functions.

\begin{table}[t]
\centering
\caption{The Performance of Error Propagation Tracking (\%)}
\label{tab:method_comparison}
\normalsize
\renewcommand{\arraystretch}{1}
\begin{tabular}{lccccc c}
\toprule
\multirow{2}{*}{\textbf{Method}} & \multicolumn{5}{c}{\textbf{Hop}} & \multirow{2}{*}{\textbf{Total}} \\
\cmidrule(lr){2-6} 
& \textbf{0} & \textbf{1} & \textbf{2} & \textbf{3} & \textbf{$\ge$4} & \\

\midrule
\midrule

\alias & \textbf{100}  & \textbf{100}  & 95.2 & \textbf{100}  & \textbf{85.7} & \textbf{97.0} \\
Static Analysis      & 100  & 90.4 & \textbf{98.8} & 72.9 & 66.1 & 90.7 \\
Internal Agent    & 100  & 87.1 & 90.5 & 84.6 & 57.1 & 87.1 \\
CoReQA        & 75 & 67.7 & 54.8 & 53.8 & 14.3    & 57.4 \\
Pure LLM       & 100 & 64.5 & 45.2 & 30.8 & 0    & 50.5 \\
\bottomrule
\end{tabular}
\end{table}

\textbf{LLM-based Methods.} We evaluate against three LLM-based approaches that differ fundamentally in their approach to accessing and reasoning about source code. 
To ensure a fair comparison, all methods (including \alias) utilize Deepseek V3 (0324)~\cite{deepseekai2024deepseekv3technicalreport} as the base model.
For each error template, all models are given the same log message and tasked with generating the error propagation path.

\begin{itemize}[noitemsep,leftmargin=5.5mm]
    \item \textit{Internal Code Agent:} This is a general-purpose Software Engineering (SWE) agent currently used in \company. Like \alias, it employs the ReAct framework~\cite{DBLP:conf/iclr/YaoZYDSN023}, but it is designed to mimic a human developer's flexible, open-ended approach to code exploration rather than being an expert system engineered for a single task.

    \item \textit{CoReQA}~\cite{chen2025coreqa}: In contrast to an iterative agent, CoReQA uses a simpler Retrieval-Augmented Generation (RAG) approach. It performs a one-shot retrieval to find and extract the text of potentially relevant functions from the codebase based on the error log. This retrieved context is then provided in a single prompt to the LLM.

    \item \textit{Pure LLM:} This is a crucial baseline to evaluate the raw reasoning capability of the LLM. The model is provided with the entire, unmodified source code of the relevant microservice in its prompt and is tasked with generating the path without any pre-filtering or iterative guidance.

\end{itemize}


\subsubsection{Evaluation Metrics}

We utilize the following metrics to evaluate the effectiveness and efficiency of different methods.


\textbf{Effectiveness Metrics.} Our primary metric for effectiveness is \textit{Accuracy}, which measures the percentage of error templates for which a method's predicted propagation path exactly matches the ground truth:

\begin{equation}
    \textit{Accuracy} = \frac{\text{The~number~of~correctly~predicted~paths}}{\text{The~total~number~of~error~templates}}
\end{equation}

Since the Static Analysis baseline outputs a set of candidate functions rather than a single path, standard accuracy is not applicable. Therefore, we evaluate its performance using \textit{Precision}. This metric measures the proportion of correctly identified functions within the full set of candidates, averaged across all templates. Let $T$ be the set of all error templates. The overall score is calculated as:


\begin{equation}
    \textit{Precision} = \frac{1}{|T|} \sum_{t \in T} \frac{|\text{ground\_truth\_path}(t)|}{|\text{candidate\_paths}(t)|}
\end{equation}

\noindent where $|\text{ground\_truth\_path}(t)|$ is the number of functions in the correct path for a given template $t$, and $|\text{candidate\_paths}(t)|$ is the total number of unique functions in the set of candidates identified for the same template.

\textbf{Efficiency Metrics.}
To evaluate practical viability, we use \textit{Inference Time} as the efficiency metric. This metric measures the average computational time (in seconds) required for a method to reconstruct the complete error propagation path for a single log template.
For iterative, agent-based methods like \alias and the Internal Code Agent, the total time is largely affected by the length of the propagation path, which determines the number of iterations. In each step, the latency is a combination of the LLM call (based on input and output token counts) and the execution time of any tools used by the agent.
In contrast, for one-shot methods like CoReQA and the Pure LLM, the time is controlled by a single, large LLM invocation. The latency is therefore mainly influenced by the size of the prompt and the length of the generated response.



\subsection{RQ1: The Effectiveness of \alias}
\label{sec:rq1_effectiveness}

The effectiveness evaluation results are presented in Table~\ref{tab:method_comparison}. With a total accuracy of 97.0\%, \alias significantly outperforms all baselines. This high level of accuracy is consistently maintained even on complex error paths with multiple hops, a scenario where other approaches begin to degrade. The subsequent analysis explores the key design choices that contribute to this robust performance.

The results first show that a candidate generation phase is essential for enabling LLMs to reason effectively in this domain. The Pure LLM baseline, which provides the model with the entire codebase as raw context, performs poorly, achieving only 50.5\% accuracy. This confirms that without a focused search space, the LLM is unable to reliably navigate the vast complexity of a microservice codebase to identify the correct error path. In contrast, \alias's performance demonstrates that the high-quality candidate paths generated by our static analysis phase are a critical prerequisite for focusing the LLM's reasoning capabilities.

Furthermore, the evaluation highlights the significant precision boost provided by the LLM-guided reconstruction phase. While our static analysis alone provides a strong set of candidates (reflected in its 90.7\% precision score), it cannot by itself disambiguate between multiple plausible paths. \alias improves this result to a final accuracy of 97.0\%, which showcases the LLM's indispensable role in analyzing, ranking, and selecting the single correct path from the statically-generated candidates.
This capability is particularly crucial in complex error scenarios where resolving ambiguity requires a deep semantic understanding of the source code.
For paths with three or more hops, the precision of static analysis degrades, whereas \alias's accuracy remains consistently high. For example, on paths of four or more hops, \alias (85.7\%) is substantially more accurate than the Static Analysis baseline (66.1\%), proving that its reasoning capability is vital for solving the long-tail and complex diagnostic challenges.

A key premise of our approach is that each log template corresponds to one error propagation path.
This is occasionally violated by the log parsing tool, Drain, when it misinterprets static keywords as variable parameters. This leads to the incorrect grouping of log events from different error paths into a single template. As a result, the ambiguous template prevents \alias from accurately reconstructing the error propagation paths, constituting the primary source of the 3\% of the failed cases in our evaluation.





\subsection{RQ2: The Efficiency of \alias}

\begin{figure}[h]
    \centering
    \vspace{-7pt}
    \includegraphics[width=0.87\linewidth]{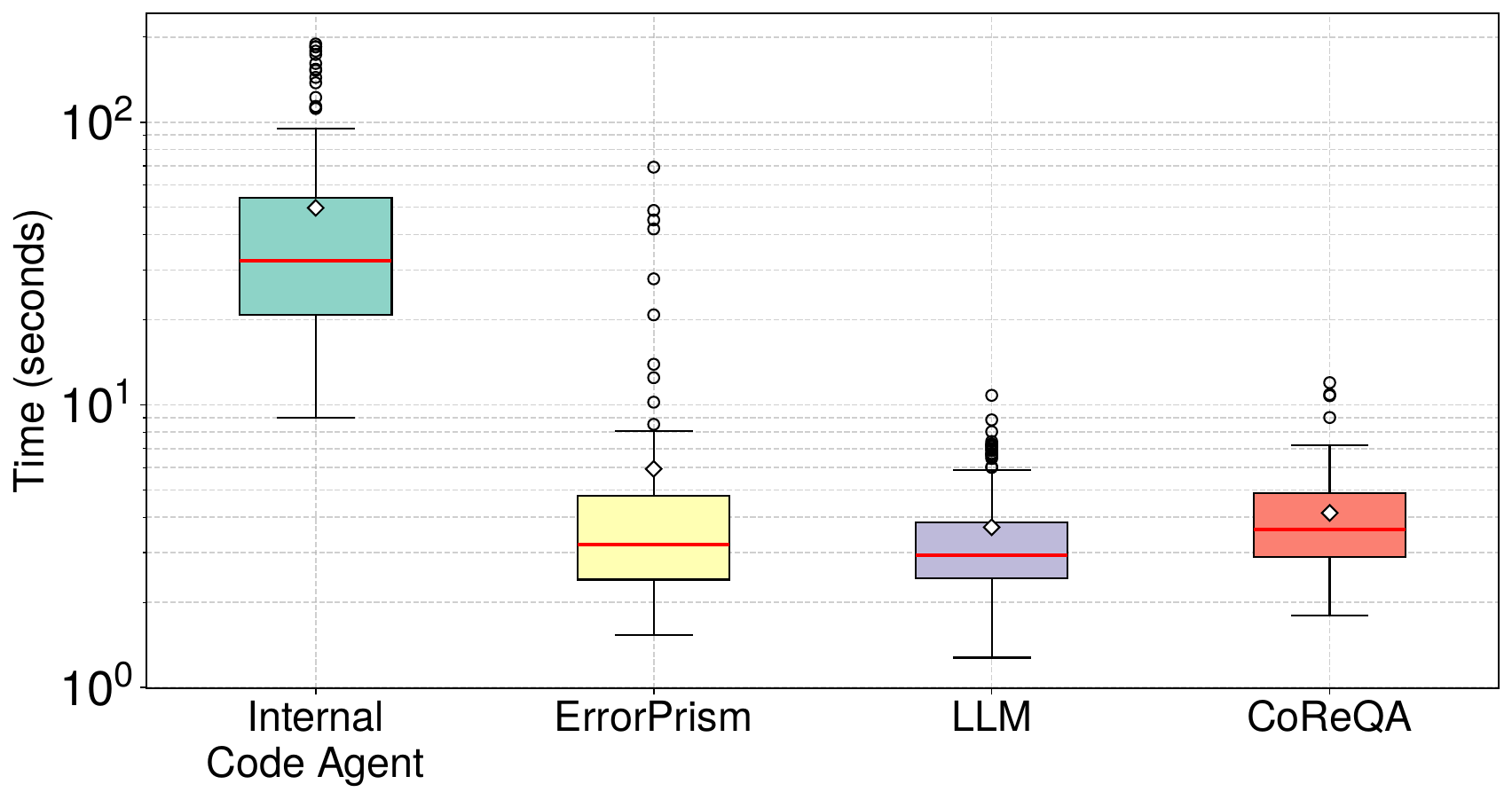}
    \caption{The Inference Time of Error Propagation Tracking}
    \label{fig:box_plot}
    \vspace{-1pt}
\end{figure}

The efficiency results shown in Fig.~\ref{fig:box_plot} highlight the practical advantages of \alias's hybrid design.
With an average inference time of \SI{5.93}{s}, \alias is approximately 8.4x faster than the Internal Code Agent (\SI{49.75}{s}).
This baseline, as a general-purpose framework, must explore a vastly larger search space. This forces it into a costly iterative loop of reasoning, tool use, and observing lengthy outputs (e.g., entire code files). As a result, its conversational history with the LLM snowballs, making each turn progressively slower and more computationally expensive. In contrast, \alias uses static analysis to massively prune the search space before engaging the LLM agent, providing it with a highly focused context.
The box plot also reveals a long tail in \alias's performance distribution, indicating that some cases have longer inference times.
These outliers typically correspond to errors with long and complex propagation paths where the initial static analysis is less effective at pruning the candidate set. In these scenarios, the agent must perform more iterative steps to traverse the ambiguous path, increasing the overall time.


Interestingly, the performance distributions of Pure LLM and CoReQA do not exhibit a similar long tail.
This can be attributed to their one-shot architecture, which makes their inference time dependent only on the size of the prompt and the response length, rather than the logical complexity of the path.
However, this architectural choice is the fundamental reason for their low accuracy.
The one-shot design means that if the initially retrieved context is incomplete or misleading, the model has no mechanism to recover or supplement the information through subsequent interactions. It must generate a final judgment in a single pass from potentially flawed input, making it highly prone to error. Therefore, while these baselines may have a lower median inference time, their efficiency is achieved at the expense of accuracy, severely limiting their practical value as discussed in RQ1.

\subsection{Case Study From Production Deployment}


To demonstrate \alias's practical effectiveness, we present a case study from its deployment in our production environment, where it solved a diagnostic challenge that is intractable for both static analysis and modern AIOps tools.

\begin{figure}[t]
    \centering
    \includegraphics[width=0.87\linewidth]{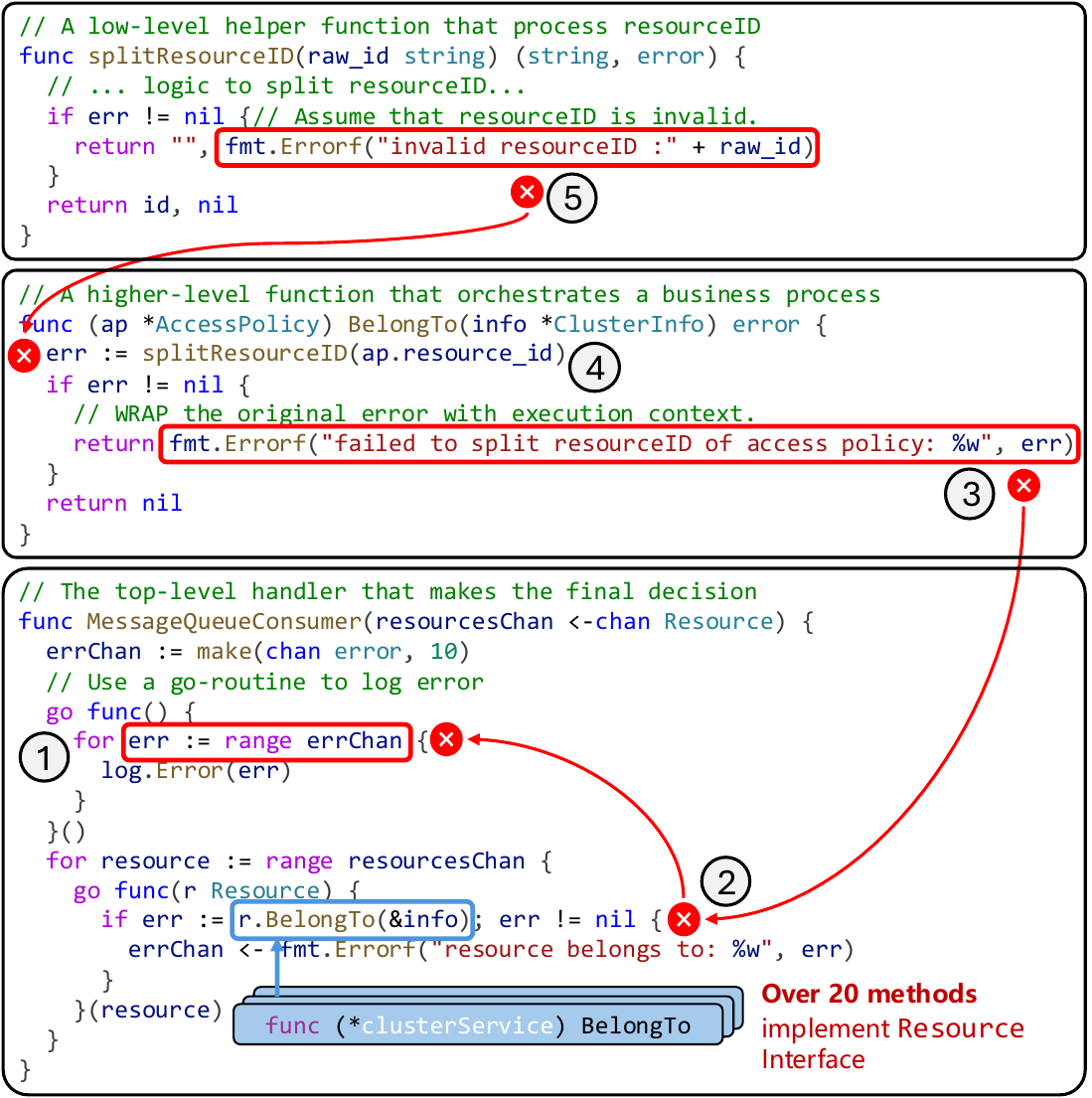}
    \caption{An Error Tracking Case in Production Systems}
    \label{fig:industrial_example}
    \vspace{-13pt}
\end{figure}

The incident began with control plane alerts, accompanied by a high volume of error logs. After considerable manual effort, engineers isolated a recurring, composite error message:


\begin{lstlisting}
resource belongs to: failed to split resourceID of access policy: invalid resourceID: Delete-123-456-cluster-prod-west-a
\end{lstlisting}

\noindent Existing code-blind tools~\cite{24WittkoppLogRCA} are ineffective in this case. While they can flag the message as an anomaly, they offer no actionable insight into the failure's origin or its multi-part structure.
This left engineers with only a high-level symptom.
The manual investigation, detailed in Fig.~\ref{fig:industrial_example}, reveals a non-trivial failure path that is challenging for conventional static analysis. The error propagates first through an interface method call \texttt{r.BelongTo} (\ding{193}) and then across an asynchronous boundary via a Go channel (\ding{192}). After a deep manual trace, the root cause was finally found in the low-level \texttt{splitResourceID} function (\ding{196}). Its implementation was built on the hard-coded assumption of a simple, four-part, hyphen-delimited string: \texttt{[action-type]-[policy-id]-[account-id]-} \texttt{[cluster-id]}. The failing ID (\texttt{Delete-123-456-} \texttt{cluster-prod-west-a}), however, violated this format because its cluster-ID part (\texttt{cluster-prod-west-a}) itself contained hyphens, which is a new naming convention for recently provisioned clusters. The legacy string-splitting logic could not handle this variation, causing the parser to fail.

\alias is able to automate the entire manual investigation process.
Its LLM-guided agent successfully reconstructs the complex propagation path by reasoning about the code's semantics.
Specifically, it 1) resolves the \texttt{r.BelongTo} interface call (\ding{194}$\rightarrow$\ding{193}) by semantically matching the error message fragments to the correct implementation, and 2) traces the \texttt{err} variable's flow through the asynchronous \texttt{errChan} (\ding{193}$\rightarrow$\ding{192}). As highlighted in Fig. \ref{fig:industrial_example}, with over 20 methods implementing the \texttt{Resource} interface, the static analyzer cannot infer the concrete runtime type of the \texttt{resource} object. Thus, it must treat each one as a potential source of the error.
By semantically matching the error message fragment \texttt{"access policy"} to the source code, \alias correctly identifies the \texttt{AccessPolicy} implementation of the \texttt{Resource} interface. This enables it to trace the \texttt{err} variable's flow through the asynchronous \texttt{errChan} and pinpoint \texttt{splitResourceID} as the origin of the failure. This delivers the exact code path that took the engineer considerable time to find.
In our workflow, this precisely identified path is provided as context to a SWE Agent for a focused, final-mile analysis. This allows the agent to deduce the root cause, i.e., the string-splitting logic's failure to handle the new cluster naming convention, providing a complete and actionable diagnosis.


\vspace{-1pt}
\subsection{Limitation Discussion}

We acknowledge some limitations of our study. First, the selection of code repositories for static analysis presents a trade-off. An overly broad scope can overwhelm the constant transitive closure and harm efficiency, while a narrow scope may miss crucial code, reducing accuracy. We mitigate it by working with developers to select a set of 67 internal microservices repositories, which contain the propagation paths for the vast majority of the error logs under study. Additionally, \alias's methodology is tailored for languages like Go that treat errors as explicit return values. While exception-based languages like Java face challenges similar to error obfuscation during execution flow recovery~\cite{DBLP:conf/ccs/Du0ZS17,DBLP:conf/asplos/YuanMXTZP10}, applying our framework to support this paradigm
requires different static analysis techniques.
We leave this for future work.


%% file: Sections/07-related.tex
\section{Related Work}
\label{sec:related}

\textbf{Fault Localization.}
Reconstructing the causal chain of an error and tracing its propagation path among services is a long-standing challenge in fault diagnosis and system debugging~\cite{sigelman2010dapper,DBLP:conf/www/YuCCGHJWSL21,DBLP:conf/infocom/ChenQZH14}.
Many methods analyze the root causes of errors by mining patterns in observability data. For instance, Minesweeper~\cite{20minesweeper} performs root cause analysis by comparing error patterns between buggy and normal traces from application telemetry. While this method is effective, it inherently relies on aggregated telemetry data, with its core lying in statistical-level pattern isolation, and does not explain the causes of errors at the code level.
Some methods use static analysis to analyze potential microservice defects. Zhang et al.~\cite{25micans} proposed a pointer-analysis-based method for constructing higher-precision cross-component call graphs in microservices, and showed these extended graphs can be applied to cross-component taint analysis. CIMET~\cite{25CIMET}, proposed by Cerny et al., uses static analysis to build inter-microservice call graphs and identify potential anti-patterns. While effective, its analysis is limited to predefined rules and ignores the program's underlying semantics.


In contrast, \alias is designed to trace multi-hop error propagation paths at the code level. By moving beyond predefined rules, it achieves this through a primarily static solution without requiring program execution.

\textbf{LLM for Program Analysis.} LLMs are transforming the field of program analysis~\cite{wang2025contemporary}. This emerging paradigm seeks to overcome the limitations of traditional, rule-based systems by applying the inherent ability of model to reason about the semantics of program, rather than relying solely on syntactic patterns.
Several recent works use LLMs to interpret the output of other analysis tools. For example, LLift~\cite{li2024enhancing} improves binary taint analysis by using an LLM to handle bug-specific modeling and navigate large codebases. A key insight from this work, which informs our design, is the finding that LLMs can reason more effectively about raw source code than about intermediate representations (IRs). The work of Chapman et al.~\cite{24ChapmanInterleaving} interleaves static analysis (EESI) with an LLM. In this approach, intermediate results from the static analysis are used to prompt the LLM for error specifications, which are then fed back into the analyzer to trace the error path.

Instead of classifying the output of analysis tools or generating static code properties, \alias employs a synergistic, iterative process. It leverages static analysis not as a calling tool, but as a powerful mechanism for massive search space reduction. By doing so, \alias can automatically reconstruct the complete, multi-hop error propagation path.

%% file: Sections/08-conclusion.tex
\section{Conclusion}
\label{sec:conclusion}

This paper tackles the problem of error obfuscation, which arises when the common practice of error wrapping creates ambiguous log messages that obscure a failure's true propagation path.
To solve this, we presented \alias, a framework that automates the reconstruction of the complete error path.
\alias first performs a comprehensive static analysis on source code to build a function call graph and index error-related strings, drastically pruning the search space. It then employs an LLM-guided agent to perform an iterative, semantic-aware search, accurately tracing the error from the final log message back to its origin.
Evaluated on 102 real-world errors in a large-scale production environment at ByteDance, \alias achieves 97.0\% accuracy, significantly outperforming both traditional static analysis and other LLM-based baselines.
By effectively transforming error logs into precise and actionable diagnostic paths, our work provides a practical solution for microservice reliability management.

